\documentclass[sigconf]{acmart}
\usepackage{graphicx}
\usepackage{amsmath}
\usepackage{booktabs}
\usepackage{algorithm}
\usepackage{algorithmic}
\usepackage{amsfonts}
\usepackage{multirow}
\usepackage{makecell}
\usepackage{subfigure}
\usepackage{color}
\usepackage{bm}
\usepackage{epstopdf}
\usepackage{url}
\usepackage[cal=cm]{mathalfa}
\usepackage{balance}
\usepackage{threeparttable}
\usepackage{lipsum}
\usepackage{enumitem}
 \usepackage{pifont}
 \usepackage{tabularx}
 \usepackage{makecell}
 \usepackage{float}

\AtBeginDocument{%
  \providecommand\BibTeX{{%
    \normalfont B\kern-0.5em{\scshape i\kern-0.25em b}\kern-0.8em\TeX}}}

\begin{document}
\title{LiveForesighter: Generating Future Information for Live-Streaming Recommendations at Kuaishou}

\renewcommand{\shorttitle}{LiveForesighter}

\author{Yucheng Lu$^\S$}
\affiliation{
  \institution{Kuaishou Technology}
  \country{luyucheng@kuaishou.com}
 }

\author{Jiangxia Cao$^{\S\star}$}
\thanks{$^\S$Equal contributions to this work}
\thanks{$^\star$Corresponding author.}
\affiliation{
  \institution{Kuaishou Technology}
  \country{caojiangxia@kuaishou.com}
 }

 \author{Kuan Xu}
\affiliation{
  \institution{Kuaishou Technology}
  \country{xukuan@kuaishou.com}
 }

 \author{Wei Cheng}
\affiliation{
  \institution{Kuaishou Technology}
  \country{chengwei07@kuaishou.com}
 }

 \author{Wei Jiang}
\affiliation{
  \institution{Kuaishou Technology}
  \country{jiangwei@kuaishou.com}
 }

 \author{Jiaming Zhang}
\affiliation{
  \institution{Kuaishou Technology}
  \country{zhangjiaming07@kuaishou.com}
 }
 
 \author{Shuang Yang}
\affiliation{
  \institution{Kuaishou Technology}
  \country{yangshuang08@kuaishou.com}
 }

  \author{Zhaojie Liu}
\affiliation{
  \institution{Kuaishou Technology}
  \country{zhaotianxing@kuaishou.com}
 }

 \author{Liyin Hong}
\affiliation{
  \institution{Kuaishou Technology}
  \country{hongliyin@kuaishou.com}
 }

\begin{abstract}
Live-streaming, as a new-generation media to connect users and authors, has attracted a lot of attention and experienced rapid growth in recent years.
Compared with the content-static short-video recommendation, the live-streaming recommendation faces more challenges in giving our users a satisfactory experience:
(1) \textit{Live-streaming content is dynamically ever-changing along time}.
(2) \textit{valuable behaviors (e.g., send digital-gift, buy products) always require users to watch for a long-time (>10 min)}.
Combining the two attributes, here raising a challenging question for live-streaming recommendation: \textit{How to discover the live-streamings that the content user is interested in at the current moment, and further a period in the future?}
To address the above question, we analyzed our live-streaming data and noticed two thought-provoking observations: 
(1) \textbf{When a live-streaming at `high-light moment'}, users will have a better experience.
(2) \textbf{Authors has `consistent style' in their live-streaming}, which allows us to predict the future information based on previous content.
Motivated by our observations, we propose the \textbf{LiveForesighter}, which aims at forecasting future live-streaming information to achieve better recommendation quality.
Specifically, we considering two types of live-streaming sequences to monitor the high-light moment and forecast live-streaming future information.
For the first observation to detect the high-light moment in the current moment, we employ the \textbf{statistic metric sequences of users’ positive behaviors} (e.g., click, buy) to reflect whether the live-streaming is in its `high-light moment' status.
For the second observation to forecast the live-streaming information in a period of time in the future, we utilize \textbf{past sold products sequence to predict future products} to enhance online-shopping live-streaming recommendation.
To our knowledge, this paper is the first work to enhance live-streaming recommendation from future information prediction perspective, and we conduct extensive offline\&online experiments and ablation analyses to demonstrate our LiveForesighter effectiveness. 
From July 2024, our LiveForesighter has been widely deployed on various services at Kuaishou, supporting 400 Million active users daily.

\end{abstract}



\maketitle

\section{Introduction}
\begin{figure*}[t!]
  \centering
\includegraphics[width=18cm,height=6cm]{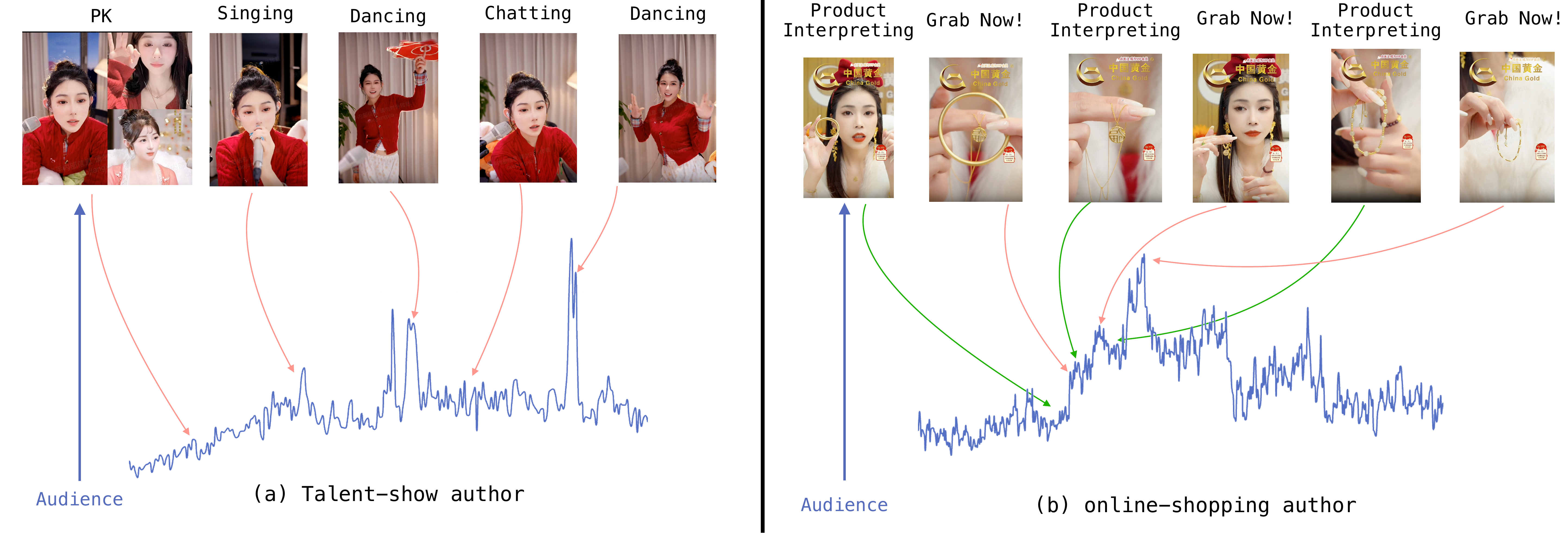}
  \caption{Live-streaming high-lights could be reflected by the crowd of users behaviours.}
  \label{case1}
\end{figure*}
\begin{figure*}[t]
  \centering
\includegraphics[width=18cm,height=4.5cm]{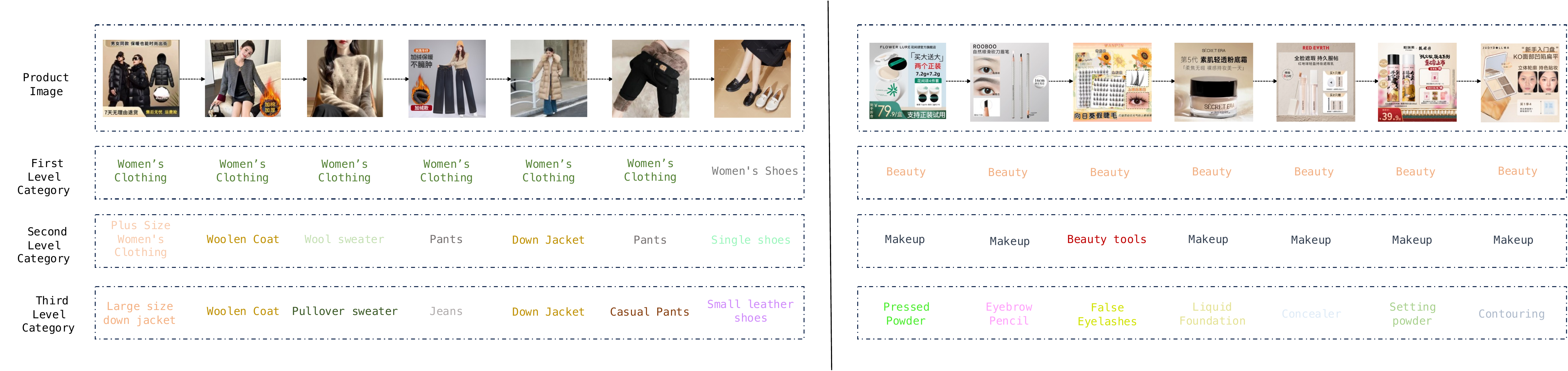}
  \caption{The interpreting product sequence of a online-shopping live-streaming.}
  \label{case2}
\end{figure*}
In recent years, the new generation short-video\&live-streaming media applications like Kuaishou, Tiktok and Xiaohongshu have attracted a surge of attention world wide and have grown rapidly.
In a broad sense, these platforms are generally play an entertainment role that users use them to watch some interesting content recommended by our algorithm.
As a result, to support our business at Kuaishou, strong and robust short-video/live-streaming recommendation systems (RecSys)~\cite{trinity, widedeep, fm} is the foundation stone to guarantee our users have satisfactory experience.

Different from the content-static short-video media, the live-streaming is more complex media to make appropriate recommendation to connect our users and authors~\cite{sliver}:
(1) \textit{Live-streaming content is dynamically ever-changing along time}, therefore users will have different experiences at different time points for a same live-streaming, then further leading different user behaviours. 
(2) Live-streaming as a service with revenue attributes, \textit{some valuable behaviors require users to watch for a long-time (>10 min)}\cite{dfm}, e.g., send digital-gift in talent-show live-streaming, buy products in online-shopping live-streaming.
To our live-streaming services, a success recommendation would follow the following data funnel chain~\cite{esmm}: \textbf{exposure} $\to$ \textbf{click} (interesting in current moment content) $\to$ \textbf{long-view/gift/buy} (interesting in a period of future content).

To this end, our live-streaming RecSys not only needs to capture the user's interests~\cite{twin, twinv2} to know which type of live-streaming he/she is interested in, but more importantly is to recommend such live-streamings at the right moment to our users. 
Specifically, the user interesting modeling technique is widely explored by many elaborate works, such as DIN~\cite{din}, ETA~\cite{eta}, etc.
Therefore, in live-streaming RecSys, the first step that we need to consider additionally is that: \textit{for a live-streaming, what moment is the best timing to distribute?}
Here we analyze our live-streaming empirical performance in Figure~\ref{case1}.
Specifically, in Figure~\ref{case1}(a), we show a talent-show live-streaming case: this type of authors will spend a lot of time chatting with audiences, competing in PK battles with other authors, and occasionally showcasing his/her talents in singing/dancing or others.
From the user feedback in terms of audience number, we could find that when the author is at dancing/singing status, the number of audiences will be significantly improved.
Besides, in Figure~\ref{case1}(b), we show another online-shopping live-streaming case: this type of authors usually sale products one by one.
For each product, they start by spending a lot of time introducing and interpreting the advantages of a product and chatting with audiences, then provide a countdown before releasing the introduced product to grab, often accompanied by discounts.
Analogously, in terms of audience number metric, we could find that when the author is introducing a product status, the number of audiences entering the room is relatively stable, but when the product is ready to grab, the number of audiences entering the room will increase suddenly.
From the two cases, we could find that different types live-streamings have different characters, and their `high-light moment' content also represents different meanings behind them.
However, we can safely draw another conclusion: \textbf{when a live-streaming at its `high-light moment', users will have a better experience and express more positive behaviours.}
As a result, we could apply the \textbf{users positive behaviours increasing trend}, to discover high-light moments adaptively for more smart recommendation.

Not only find the best live-streaming timing to distribute, another goal of our RecSys is to convert users to gift author or buy some products, which requires our users should be deeply interested in the live-streaming to watch a long-time.
As a result, in live-streaming RecSys, the second step that we need to consider further is: \textit{in addition to the current moment, does the future content also meet users interests?}
Here we give our online-shopping live-streaming as example in Figure~\ref{case2}.
Specifically, we shown two online-shopping authors live-streaming sold products and corresponding categories sequences in coarse-to-fine three-level chronological order.
The left-side author mainly sells women's clothing, but the categories are quite diverse (e.g., clothes, pants, shoes, etc.), and the second level and three level categories are constantly changing.
The right-side author mainly sells some beauty products, but the products are relatively vertical and mainly the third-level categories are changing.
Fortunately, although the products are various, we could find that the highly related products in their vertical filed are interpreted/sold consecution, e.g., Jacket $\leftrightarrows$ Coat $\leftrightarrows$ Sweater, Liquid Foundation $\leftrightarrows$ Concealer $\leftrightarrows$ Setting Powder.
From the two cases, we could observe that \textbf{Authors has ‘consistent style’ in their live-streaming}, which allows us to predict the \textbf{future later products have some correlation with the previous products}.

In this paper, we present our simple-yet-effective generative model, \textbf{LiveForesighter}, a new paradigm for live-streaming modeling to achieve better recommendation quality.
Specifically, we consider two types of live-streaming sequences to monitor the high-light moment and generate live-streaming future information.
For the first challenge to mine the live-streaming high-light moment timing to distribution, as discussed above, our key insight is to identify which live-streaming is experiencing the users' positive behaviours increasing trends.
Hence, to describe such positive behaviours increasing trends, we first utilize the past live-streaming-side statistic features of users' positive behaviours sequences (e.g., click, purchase) to predict future positive behaviours trends.
Therefore, our model could not only monitor the real-time content quality to detect whether the live-streaming is in its positive behaviours increasing trends and high-light moment, but also perceive future trends to ensure that live-streaming keeps high quality in a period of time.
For the second challenge to guarantee future content should also meet users interests, in this paper, we conduct our LiveForesighter to enhance online-shopping live-streaming recommendation, and our key insight is to utilize a series of previously sold products to predict future products.
In this way, our model not only determines whether users currently enjoy the live-streaming, but also predicts whether they will continue to like its content in the near future.
In summary, our contributions are as follows:
\begin{itemize}[leftmargin=*,align=left]
\item We present in-depth analyses to show the challenges in live-streaming recommendation,
to our knowledge, this paper is the first work to enhance live-streaming recommendation from future information prediction perspective, which will shed light on other researchers to explore more robust live-streaming RecSys.
\item We develop the LiveForesighter, which considers the live-streaming side audience's positive behaviours and previous content sequences to monitor the live-streaming behaviours increasing trends and generate future content information. 
\item We conduct extensive experiments on the first-hand live-streaming scenarios at Kuaishou. Results verify the effectiveness of our LiveForesighter, which has been deployed at Kuaishou, contributing significant improvement to our platform.
\end{itemize}

\section{Methodology}
This section introduces the details of our generative model~\cite{gpt}, LiveForesighter.
We first specify the common background of industrial live-streaming recommendation systems formulation~\cite{youtube, youtubemmoe}.
Next, we dive into the details of the statistic sequences modeling to monitor the real-time content quality.
Finally, we describe how we apply the previously sold products to predict future products.

\subsection{Background of Industrial RecSys}
In general, the RecSys aims at generating a small dozen item set that maximizes user interests from their large amount of implicit feedbacks~\cite{recflow}.
Nevertheless, in industrial settings, accurately and effectively searching such dozen items from hundreds of millions of item pool is no easy task. 
To make a trade-off between the cost and precise, in recent years, a widely adopted solution by major companies is a two-stage paradigm: candidate generation~\cite{tdm, kuaiformer} and ranking~\cite{ple, home}:
\begin{itemize}[leftmargin=*,align=left]
\item For candidate generation stage, which aims at utilizing multiple sources, signals and models to find users interested hundreds of item candidates from million-scale item pool~\cite{ebr}. For example, identifying high-quality authors who are within categories that align with user preferences.
\item For ranking stage, which aims at finding the best dozen items from the retrieved hundreds of item candidates~\cite{pepnet}. For example, finding those live-streamings which in high-light moment or positive behaviours increasing trends status from all candidates.
\end{itemize}
Obviously, the candidate generation stage (e.g., user interests modeling) and ranking stage (e.g., user-item pair modeling) play very different roles in the RecSys chain, therefore the used techniques are greatly different.
For example, candidate generation focuses on item-item relation mining~\cite{swing}, user interests compression~\cite{kuaiformer}, and the ranking stage focuses on user-item cross feature mining~\cite{fm,sim}, multi-task learning~\cite{MTLSurvey}.
In this paper, we focus on extending the later \textbf{ranking stage}, towards to build more comprehensive live-streaming RecSys.

\begin{figure*}[t!]
  \centering
\includegraphics[width=18cm,height=8cm]{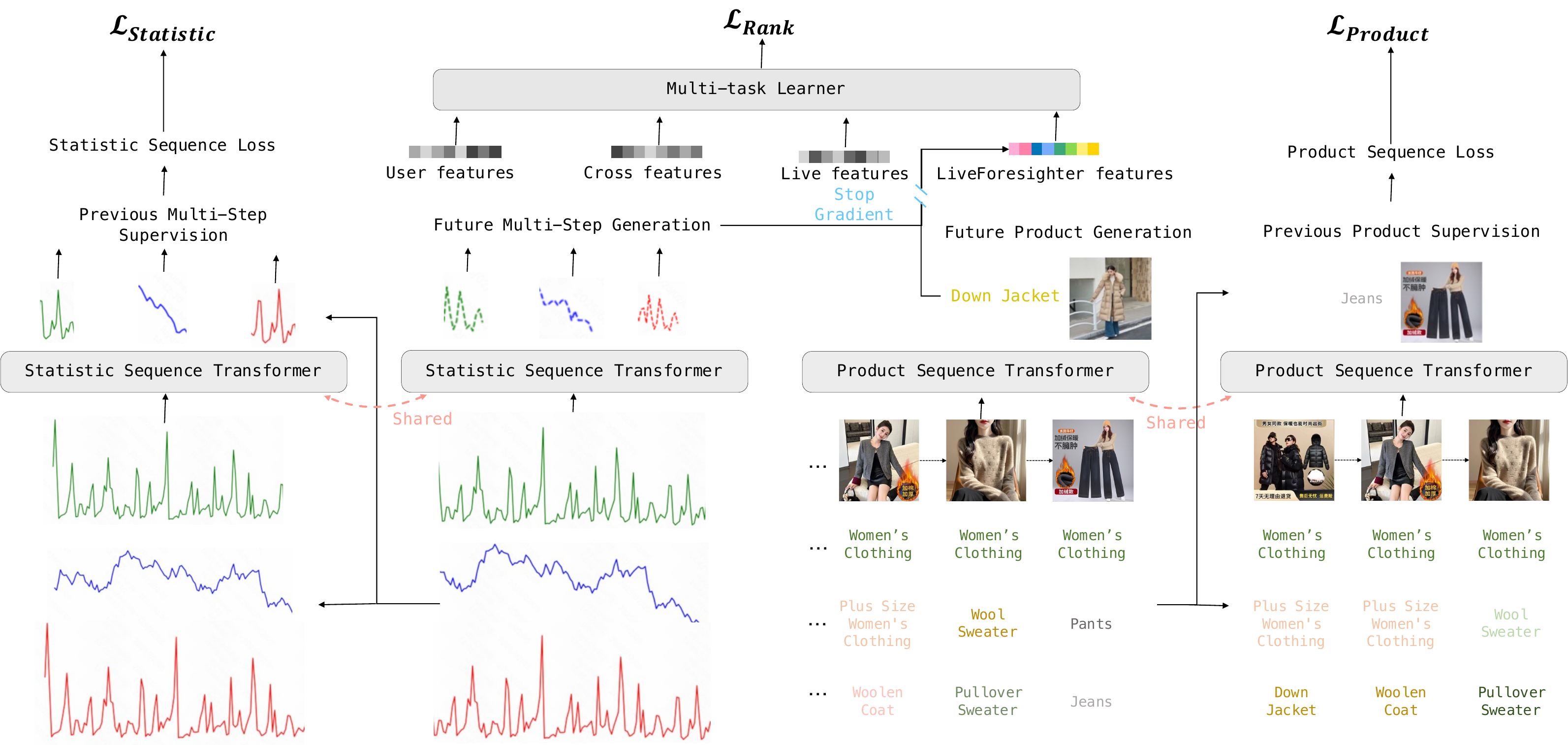}
  \caption{An overview of model architecture of LiveForesighter.}
  \label{model}
\end{figure*}

\subsubsection{Base Ranking Model Architecture}
We first briefly retrospect our base ranking model design.
In practice, the goal of ranking model is to predict the certain probabilities (e.g., click rate \texttt{ctr}, long-view rate \texttt{lvtr}, purchase rate \texttt{cvr}), to measure the recommendation value for a user-item pair.
Specifically, in the ranking model training, each user-item sample is formed as a series of discretization categorical features, such as user ID, user gender, item ID, item category ID, etc, and cross sequence features like user’s previous clicked sequence within same item category ID.
To simplify, we denote each training sample feature and labels as $[\mathbf{V}_\texttt{user}, \mathbf{V}_\texttt{live}, \mathbf{V}_\texttt{cross}]$ and $\{y^\texttt{ctr}, y^\texttt{lvtr}, \dots\}\in \{0,1\}$.
Based on the user-item pair input information and its collected labels, the ranking model $\texttt{Rank}(\cdot)$ training process is formulated as:
\begin{equation}
\begin{split}
\widehat{y}^\texttt{ctr}, \widehat{y}^\texttt{lvtr}, \dots = &\texttt{Rank}\big([\mathbf{V}_\texttt{user}, \mathbf{V}_\texttt{live}, \mathbf{V}_\texttt{cross}]\big),\\
\mathcal{L}_{\texttt{rank}} = - \sum_{\texttt{xtr}}^{\texttt{ctr}, \texttt{lvtr}, \dots} \big(y^{\texttt{xtr}}\log &\widehat{y}^{\texttt{xtr}} + (1-y^{\texttt{xtr}})\log (1-\widehat{y}^{\texttt{xtr}})\big)\\
\end{split}
\label{base}
\end{equation}
where $\texttt{Rank}(\cdot)$ is an MLP based network, $\widehat{y}^\cdot$ means model predict score, $\mathcal{L}_{\texttt{rank}}$ is the ranking loss function that consists of multiple binary cross-entropy loss.
In this paper, \textbf{our LiveForesighter focus on generating the representation of live-streaming information $\mathbf{V}_\texttt{livefore}$} to enhance ranking model ability.

\subsection{Statistic Sequence Modeling}
Although the wide-used live-streaming side features (i.e., author ID, live-streaming category ID, etc) could represent the authors information to some extent.
However, it is still hard to describe real-time changed live-streaming content and answer the challenge: \textit{for a live-streaming, what moment is the best timing to distribute?}
Fortunately, we observe that these high-light moment timing could be discovered via the users positive behaviours increasing trends adaptively.
Thereby we first explicitly introduce some statistic sequences to inject such behaviours trends knowledge to empower the live-streaming side information.
In practice, we add the following categories of statistic sequences to empower our LiveForesighter:
\begin{itemize}[leftmargin=*,align=left]
\item Out-Room live-streaming information: Exposure, Audience, etc.
\item Convert information: Gmv value, Order, Gift value, etc.
\item Interaction information: Comment, Like, etc.
\item In-Room live-streaming information: Click product/cart, etc.
\end{itemize}
In practice, we collected these sequences every 30s, and form them as $\mathbf{S} = \{\mathbf{s}_1, \dots, \mathbf{s}_N\}\in \mathbb{R}^{N\times T}$ to represent $N$ sequences with $T$ time steps.
Specifically, each sequence $\mathbf{s}_i = [s_{i,1}, \dots, s_{i,T}]$ and $s_{i,j}$ is an integer to represents the real collected statistic value.
For convenience, we utilize the $\mathbf{S}_{1:T-5}\in \mathbb{R}^{N\times (T-5)}$ to denote the whole sequences that time step indexed in the range $[1, T-5]$.
%

Inspired by the self-supervised sequence generation~\cite{sutskever2014sequence}, we first conduct the sequence generation to encourage our model to capture the sequences transform pattern as follows (\textbf{here we assume to predict next 5 steps}):
\begin{equation}
    \begin{split}
        \bar{\mathbf{S}}_{1:T-5}, \bm{\mu}, &\bm{\delta} = \texttt{ReVIN}(\mathbf{S}_{1:T-5}),\\
        \bar{\mathbf{S}}_{T-4:T}, \mathbf{E}_{1:T-5}^{\texttt{Stat}} = \texttt{St}&\texttt{atisticTransformer}(\bar{\mathbf{S}}_{1:T-5}),\\
        \widehat{\mathbf{S}}_{T-4:T} = &\texttt{De-ReVIN}(\bar{\mathbf{S}}_{T-4:T}, \bm{\mu}, \bm{\delta}),\\
        \mathcal{L}_{\texttt{Statistic}} &= \texttt{MS}\texttt{E}(\widehat{\mathbf{S}}_{T-4:T}, \mathbf{S}_{T-4:T}).
    \end{split}
    \label{statistrain}
\end{equation}
where the $\texttt{ReVIN}(\cdot)$ and $\texttt{De-ReVIN}(\cdot)$ is the sequence normlization~\cite{kim2021reversible} technique (i.e., $\bm{\mu}\in \mathbb{R}^N$, $\bm{\delta}\in \mathbb{R}^N$), the $\mathbf{E}_{1:T-5}\in \mathbb{R}^{N\times D}$ is the input sequence encoding results, the $\widehat{\mathbf{S}}_{T-4:T}\in \mathbb{R}^{N\times 5}$ denotes the multiple step prediction results, the $\texttt{MSE}(\cdot)$ denote the training loss to reflect the prediction accuracy, and the $\texttt{StatisTransformer}(\cdot)$ denotes an arbitrary Transformer-style sequences encoding method.
Indeed, the statistic sequence generation is closely related to the time forecasting filed problem definition, and there are many elaborate works such as PatchTST~\cite{nie2022time}, iTransform~\cite{liu2023itransformer}, etc.
In our LiveForesighter, for simplicity, we utilize the iTransform to implement our $\texttt{StatisTransformer}(\cdot)$ for sequences generation.

By optimizing the training objective $\mathcal{L}_{\texttt{Stat}}$, our model could have the ability to predict future behaviours trends for different live-streamings.
Therefore, in real-time recommendation, we could utilize the full current behaviors trends to predict a period of future information:
\begin{equation}
    \begin{split}
        \bar{\mathbf{S}}_{5:T}, \bm{\mu}, &\bm{\delta} = \texttt{ReVIN}(\mathbf{S}_{5:T}),\\
        \bar{\mathbf{S}}_{T:T+5}, \mathbf{E}_{5:T}^{\texttt{Stat}} = \texttt{St}&\texttt{atisticTransformer}(\bar{\mathbf{S}}_{5:T}),\\
        \widehat{\mathbf{S}}_{T:T+5} = &\texttt{De-ReVIN}(\bar{\mathbf{S}}_{T:T+5}, \bm{\mu}, \bm{\delta}),\\
        \mathbf{V}_{\texttt{livefore}}\longleftarrow \texttt{stop}&(\texttt{flatten}(\widehat{\mathbf{S}}_{T:T+5}), \texttt{flatten}(\mathbf{E}_{5:T}^{\texttt{Stat}}))
    \end{split}
    \label{statisinfer}
\end{equation}
where the $\widehat{\mathbf{S}}_{T:T+5}$ means the future prediction user behaviors trends, $\texttt{flatten}(\cdot)$ is a re-size operator, $\texttt{stop}(\cdot)$ is the stop gradient operator to detach the downstream recommendation influence.
In this way, equipped with the current sequence representation $\mathbf{E}_{1:T}^{\texttt{Stat}}$ and prediction results $\bar{\mathbf{S}}_{T:T+5}$, our ranking model could not only monitor the real-time content quality to detect whether the live-streaming is in its positive behaviours increasing trends and high-light moment, but also perceive future trends to ensure that live-streaming keeps high quality in a period of time.

\subsection{Products Sequence Modeling}
Instead of finding high-quality live-streaming for our users, another goal of our RecSys is to convert users to gift author or buy some products, which requires our users should be deeply interested in the live-streaming to watch a long-time.
This section aims at answering another challenge: \textit{in addition to the current moment, does the future content also meet users interests?}
Actually, live-streaming author has a `consistent style' in their live-streaming, therefore we could utilize the previous live-streaming content to forecast future information.
Here we focus on predicting future products information to enhance online-shopping live-streaming recommendation.

In practice, we could collect a live-streaming's previous sold products ID and corresponding category information.
We form the product sequence as $\mathbf{I} = \{[p_1, c_1^1, c_1^2, c_1^3],\dots,[p_T, c_T^1, c_T^2, c_T^3]\}\in \mathbb{R}^{T\times4}$, where the $p_i$ denote the product ID, $c^1/c^2/c^3$ are the category information from coarsen-to-fine.
Similarly, we also extend the self-supervised idea to empower our LiveForesighter to have the ability to generate future product information.
Different from the statistic sequences multi-step prediction, due to the time cost and huge space for generating the next product ID, thus we only generate the finest granular category information of the next product, to make a trade-off between the effects and efficiency:
\begin{equation}
    \begin{split}
        \bar{\mathbf{I}}_{1:T-1} = \texttt{Embedding}&\texttt{Lookup}(\mathbf{I}_{1:T-1}, \mathbf{P}, \mathbf{C}^1, \mathbf{C}^2, \mathbf{C}^3),\\
        \widehat{\mathbf{c}}_T^3, \mathbf{E}_{1:T-1}^{\texttt{Prod}} = \texttt{Pro}&\texttt{ductTransformer}(\bar{\mathbf{I}}_{1:T-1}),\\
        \mathcal{L}_{\texttt{Product}} = &\texttt{Softmax}(\widehat{\mathbf{c}}_T^3, c_T^3),\\
    \end{split}
    \label{statistrain}
\end{equation}
where the $\mathbf{P} \in \mathbb{R}^{|p|\times D}$, $\mathbf{C}^1 \in \mathbb{R}^{50\times D}$, $\mathbf{C}^2 \in \mathbb{R}^{1000\times D}$, $\mathbf{C}^3 \in \mathbb{R}^{5000\times D}$ are trainable product and categories embedding matrices, $|p|$ denotes the amount of total products,
$\bar{\mathbf{I}}_{1:T-1} \in \mathbb{R}^{(T-1)\times 4D}$ is the selected embedding input for downstream Transformer, the $\mathbf{E}_{1:T-1}\in \mathbb{R}^{(T-1)\times D}$ means the encoding results of input sequence, the $\widehat{\mathbf{c}}_T^3 \in \mathbb{R}^{5000}$ is the generated the fine-grained category prediction distribution, the $c_T^3 \in \mathbb{R}$ denotes the training ground-truth observed labels.
For the sake of simplicity, we utilize the naive Transformer~\cite{vaswani2017attention} as our \texttt{ProductTransformer} backbone.

By optimizing the training objective $\mathcal{L}_{\texttt{Product}}$, our LiveForesighter is able to capture the potential relationship between products.
Next, we employ the latest product information to predict future product category information to support our ranking model:
\begin{equation}
    \begin{split}
        \bar{\mathbf{I}}_{2:T} = \texttt{Embedding}&\texttt{Lookup}(\mathbf{I}_{2:T}, \mathbf{P}, \mathbf{C}^1, \mathbf{C}^2, \mathbf{C}^3),\\
        \widehat{\mathbf{c}}_{T+1}^3, \mathbf{E}_{2:T}^{\texttt{Prod}} = \texttt{Pro}&\texttt{ductTransformer}(\bar{\mathbf{I}}_{2:T}),\\
        \mathbf{V}_{\texttt{livefore}}\longleftarrow \texttt{stop}(\widehat{\mathbf{c}}_{T+1}^3)&\cdot \widetilde{\mathbf{C}}^3 \oplus \texttt{stop}(\texttt{flatten}(\mathbf{E}_{2:T}^{\texttt{Prod}}))
    \end{split}
    \label{statistrain}
\end{equation}
where the $\widehat{\mathbf{c}}_{T+1}^3 \in \mathbb{R}^{5000}$ is the generated the future next product category information, and the $\widetilde{\mathbf{C}}^3\in \mathbb{R}^{5000\times D}$ is a ranking model trainable parameter matrix.
In this way, equipped with the current sequence representation $\mathbf{E}_{1:T}^{\texttt{Prod}}$ and prediction next product information $\widehat{\mathbf{c}}_{T+1}^3$, our ranking model not only determine whether users currently enjoy the live-streaming, but also predict whether they will continue to like its content in the near future.
The overall framework of our LiveForesighter is shown in Figure~\ref{model}.

\begin{figure}[t!]
  \centering
\includegraphics[width=8cm,height=4cm]{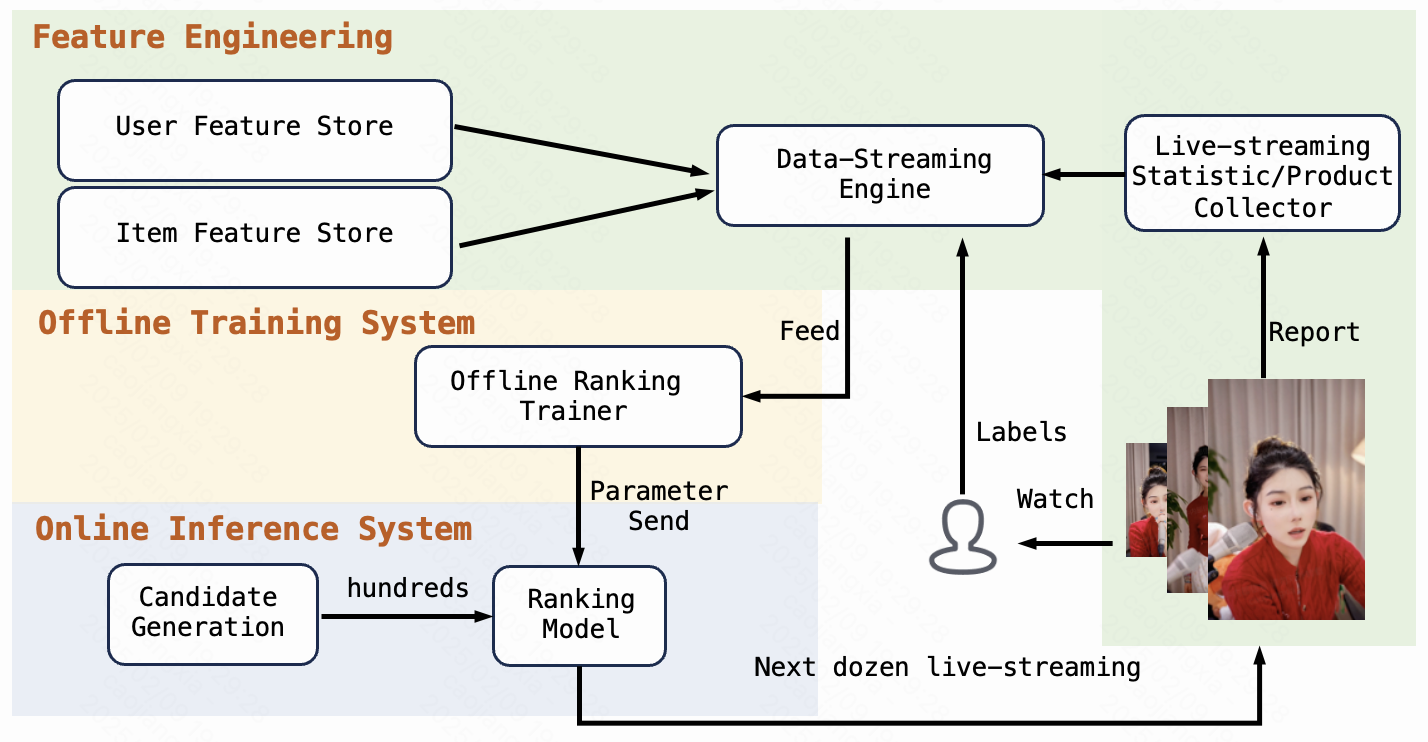}
  \caption{LiveForesighter Deployment.}
  \label{deployment}
\end{figure}

\section{LiveForesighter Deployment}
In this section, we will introduce the comprehensive deployment architecture of LiveForesighter in Figure~\ref{deployment}. The deployment architecture includes a data-streaming engine, an offline training platform and an online inference platform. The live-streaming data-streaming engine generates training data after a user watches the Kuaishou Live-streaming.
In this process, the data-streaming needs to assemble the labels, the live-streaming ID features, user ID features, and the additional statistical sequence and product sequences.
After the training samples are prepared, the data-streaming is the input to train our LiveForesighter and Ranking model, which are implemented by Tensorflow. 
As for online inference module, the training platform pushes the trained checkpoint to the online inference system after a few steps of updates in a real-time manner, to respond to the user's recommendation requests.

\begin{table*}[t!]
\centering
\caption{Offline results(\%) in term AUC, UAUC and GAUC on online-shopping live-streaming services at Kuaishou.}
\setlength{\tabcolsep}{18.5pt}{
\begin{tabular}{l|cccccc}
\toprule
\multirow{4}{*}{\makecell{LiveForesighter\\Model\\Variants}} 
& \multicolumn{6}{c}{Online-Shopping Live-Streaming Services}   \\ 
\cmidrule(r){2-7} & \multicolumn{3}{c}{CTR}  & \multicolumn{3}{c}{CVR}   \\ 
\cmidrule(r){2-4} \cmidrule(r){5-7} & AUC & UAUC & GAUC & AUC & UAUC & GAUC \\
\hline
Baseline Model & 84.01 & 62.02 & 62.06 & 83.69 & 58.30 & 58.40 \\ 
\midrule
+ Statistic Sequences & +0.093 & +0.363 & +0.349 & +0.043 & +0.092 & +0.220 \\
+ Product Sequences & +0.033 & +0.167 & +0.165 & +0.062 & +0.030 & +0.006 \\
+ Both Sequences & \textbf{+0.109} & \textbf{+0.593} & \textbf{+0.593} & \textbf{+0.102} & \textbf{+0.235} & \textbf{+0.234} \\
\bottomrule
\end{tabular}
}
\label{mainofflineshopping}
\end{table*}

\begin{table*}[t!]
\centering
\caption{Offline results(\%) in term AUC, UAUC and GAUC on talent-show live-streaming services at Kuaishou.}
\resizebox{\linewidth}{!}{
\begin{tabular}{cccccccccccccccc}
\toprule
\multirow{4}{*}{\makecell{LiveForesighter\\Model\\Variants}} & \multicolumn{15}{c}{Talent-Show Live-Streaming Services}   \\ 
\cmidrule(r){2-16}& \multicolumn{3}{c}{CTR}  & \multicolumn{3}{c}{EVTR} & \multicolumn{3}{c}{LVTR} & \multicolumn{3}{c}{CMTR} & \multicolumn{3}{c}{GTR}
\\ \cmidrule(r){2-4} \cmidrule(r){5-7} \cmidrule(r){8-10} \cmidrule(r){11-13} \cmidrule(r){14-16} & AUC & UAUC & GAUC & AUC & UAUC & GAUC & AUC & UAUC & GAUC & AUC & UAUC & GAUC & AUC & UAUC & GAUC\\
\midrule
Baseline Model & 81.565 & 64.802 & 66.125  & 80.284 & 67.081 & 68.520 & 86.159 & 72.267 & 74.291 & 93.589 & 75.108 & 76.169 & 96.887 & 72.590 & 73.198 \\
\midrule
+ Statistic Sequences & +\textbf{0.096} & +\textbf{0.231} & +\textbf{0.238}  & +\textbf{0.103} & +\textbf{0.203} & +\textbf{0.200} & +\textbf{0.156} & +\textbf{0.276} & +\textbf{0.279}  & +\textbf{0.076} & +\textbf{0.532} & +\textbf{0.521} & +0.038 & +\textbf{0.333} & +\textbf{0.336} \\
\bottomrule
\end{tabular}
}
There is only a statistical sequence because of talent-show  live-streaming does not containing product sequence signal.
\label{mainofflinetalent}
\end{table*}

\begin{table*}[ht]
\centering
\setlength{\tabcolsep}{5pt}
\caption{A/B testing results of online-shopping live-streaming services at Kuaishou.}
\begin{tabular}{cccccccccc}
\toprule
\multirow{2}{*}{Model Variants} & \multirow{2}{*}{\makecell{Gross Merchandise \\ Volume}} & \multirow{2}{*}{\makecell{Order \\ Times}} & \multirow{2}{*}{\makecell{Gross Merchandise 
 \\ Volume Per Mile}} & \multirow{2}{*}{\makecell{Total Live-streaming \\ Watching Time}} & \multirow{2}{*}{\makecell{CTR}} & \multirow{2}{*}{\makecell{CVR}} & \multirow{2}{*}{\makecell{Follow \\ Counts}} \\
\\
\midrule
\makecell{+ Statistic Sequences}& +\textbf{0.864}\% & +\textbf{1.074}\% & +\textbf{0.888}\% & +\textbf{0.681}\% & +\textbf{0.327}\% & +\textbf{0.784}\% & +\textbf{0.590}\% \\
\midrule
\makecell{+ Product Sequences}& +\textbf{0.644}\% & +\textbf{0.773}\% & +\textbf{0.464}\% & +\textbf{0.574}\% & +\textbf{0.061}\% & +\textbf{0.531}\% & +\textbf{0.355}\% \\
\bottomrule
\end{tabular}
\label{mainonlineshopping}
\end{table*}

\begin{table*}[ht]
\centering
\setlength{\tabcolsep}{10pt}
\caption{A/B testing results of talent-show live-streaming services at Kuaishou.}
\begin{tabular}{ccccccccc}
\toprule
\multirow{2}{*}{Model Variants} & \multirow{2}{*}{\makecell{Live-streaming\\Infiltration}} & \multirow{2}{*}{\makecell{Total Live-streaming\\Watching Time}} & \multirow{2}{*}{\makecell{Long-View\\Times}} & \multirow{2}{*}{\makecell{Long-View\\Users}} & \multirow{2}{*}{\makecell{Click\\Times}} & \multirow{2}{*}{\makecell{Gift Users}} \\
\\
\midrule
\makecell{+ Statistic Sequences}& +\textbf{0.149}\% &+\textbf{0.235}\% &+\textbf{0.547}\% &+\textbf{0.572}\% &+\textbf{0.290}\% &+0.138\% \\
\bottomrule
\end{tabular}
\label{mainonlinetalent}
\end{table*}

\section{Experiments}
In this section, we perform detailed offline\&online experiments on two live-streaming services at Kuaishou, to verify the effectiveness of the LiveForesighter learning paradigm. 
In general, we aim to answer the following research questions:

\begin{itemize}[leftmargin=*,align=left]
\item \textbf{RQ1}: How does LiveForesighter perform in different live-streaming services and achieve satisfactory improvements?
\item \textbf{RQ2}: How does LiveForesighter contribute online gains to our business?
\item \textbf{RQ3}: How does the future information generative accuracy influence the model performance?
\item \textbf{RQ4}: How does the statistic sequence parameters influence the model performance?
\item \textbf{RQ5}: How does LiveForesighter affect the online distribution?
\end{itemize}

\subsection{Experiments Setting}
To verify our LiveForesighter effectiveness, we conduct the first-hand large-scale validation in two live-streaming scenarios at Kuaishou, the online-shopping and talent-show live-streaming services.
It is worth mentioning that our baseline Ranking models of the two services are huge MLP-based neural-networks (about 0.1 Billion parameters) that integrates many different techniques, e.g., DIN~\cite{din}, FM~\cite{fm}, SIM~\cite{sim}, MoE~\cite{esmm}, IPW~\cite{ipw} and so on.
For the evaluation metrics, we adopt the industrial wide-used AUC (ROC version), UAUC (AUC average weighted by each user) and GAUC (AUC weighted by the exposures of each user) to show our model performance.
Considering the two different business role of these two types of live-steaming, e.g., online-shopping live-streaming for revenue, the talent-show live-streaming services for both revenue and user experience, hence we report different core tasks performance for different services, e.g., click-rate (\textbf{CTR}), convert-rate (\textbf{CVR}) for online-shopping live-streaming, and the click-rate (\textbf{CTR}), effective-view-rate (\textbf{EVTR}), long-view-rate (\textbf{LVTR}), comment-rate (\textbf{CMTR}) and gift-rate (\textbf{GTR}) for talent-show live-streaming services.

\subsection{Offline Performance (RQ1)}
The Table~\ref{mainofflineshopping} and \ref{mainofflinetalent} are shown the real-world model performances at online-shopping and talent-show live-streamings.
Considering the large-scale 400 Million user group in our application daily, it is worth noting that the stable improvements of 0.05\% in AUC, UAUC or GAUC offline are significant enough to contribute online revenue gains for our services.
According to them, we have the following conclusions:
(1) Compared with the base ranking model, directly incorporating the predicted statistic sequences could further enhance the models prediction accuracy in both of live-streaming services (+0.09\% and +0.10\% in CTR-AUC at online-shopping/talent-show live-streaming, respectively).
The experiment results indicate that utilizing the users' positive behaviours trends to monitor/generate the current/future live-streaming content quality is an effective way to find appropriate timing to distribute live-streamings.
(2) For the product sequences, since the talent-show live-streaming does not contain such product sequence signal, we only show results at online-shopping live-streaming services.
According to Table~\ref{mainofflineshopping}, we could find that incorporating the product sequences and both sequences could further enhance model performance, which demonstrates that forecasting the future information is effective for live-streaming recommendation.
(3) In our implementation, the LiveForesighter is an individual model that gradient-disentangled with the ranking model, which indicates that monitor/generate different live-streaming information from different aspects is vital for live-streaming recommendation.

\begin{table}[t]
\centering
\caption{Statistic Forecasting Accuracy Analyzing.}
\setlength{\tabcolsep}{6pt}{

\begin{tabular}{c|cccc}
\toprule
Technique & MSE & CTR-AUC & CVR-AUC \\
\midrule
Mean Value& 59 & +0.05\% & +0.03\% \\ 
Latest Value & 64 & +0.02\% & +0.02\% \\
LiveForesighter & 56 & +0.08\% & +0.04\% \\
\bottomrule
\end{tabular}
}
\label{acc_abl1}
\end{table}

\begin{table}[t]
\centering
\caption{Product Forecasting Accuracy Analyzing.}
\setlength{\tabcolsep}{6pt}{

\begin{tabular}{c|cccc}
\toprule
Technique & HitRate & CTR-AUC & CVR-AUC \\
\midrule
Most Category& 4.5\% & +0.02\% & -0.03\% \\
Latest Category & 24.1\% & +0.02\% & +0.02\% \\
LiveForesighter & 29.3\% & +0.03\% & +0.06\% \\
\bottomrule
\end{tabular}
}
\label{acc_abl2}
\end{table}

\begin{table}[t]
\centering
\caption{Statistic Sequence Importance Analyzing.}
\setlength{\tabcolsep}{3pt}{
\begin{tabular}{c|cccc}
\toprule
\multirow{2}{*}{Metrics}  & \multicolumn{4}{c}{Statistic Sequence Types} \\
\cmidrule(r){2-5}  
 & Out-Live &Convert & Interaction & In-Live\\
\midrule
CTR-AUC & +0.02\% & +0.15\% & +0.01\% & +0.01\% \\
CVR-AUC & -0.02\% & +0.01\% & +0.19\% & +0.01\% \\
\bottomrule
\end{tabular}
}
\label{importance_abl}
\end{table}

\begin{table}[t]
\centering
\caption{Statistic Forecasting Steps Analyzing.}
\setlength{\tabcolsep}{17pt}{
\begin{tabular}{c|ccc}
\toprule
\multirow{2}{*}{Metrics}  & \multicolumn{3}{c}{Generative Steps} \\
\cmidrule(r){2-4}  
 &Step-2 & Step-3 & Step-4\\
\midrule
MSE & +7.9\% & +11.4\% & +13.3\%  \\ 
CTR-AUC & +0.07\% & +0.03\% & -0.01\%  \\
\bottomrule
\end{tabular}
}
\label{step_abl}
\end{table}

\subsection{Online A/B Tests (RQ2)}
To validate the real business contribution that LiveForesighter brings to our live-streaming service, we conducted online A/B testing experiments to serve as ranking stage model.
We evaluate model performance for different services, we give the core metrics (e.g., the Gross Merchandise Volume (GMV) and Order Times for online-shopping, and User Group Live-streaming Infiltration and Total Watching Time for talent-show) and other limitation metrics (e.g., Follow and Long-View Users).
From the Table~\ref{mainonlineshopping} and Table~\ref{mainonlinetalent}, our proposed LiveForesighter variants achieve significant improvements of +0.864\%, +0.644\% in GMV and +0.149\% in infiltration, respectively.
Additionally, we find that LiveForesighter achieves satisfied improvements in terms of the online CTR/CVR, especially in the online-CVR, which reveals generating future information could alleviate the feedback delay problem.
Moreover, we can find that the long-view times and long-view users also achieve a larger improvement, which demonstrates our LiveForesighter could help system to distribute live-streaming at its high-quality moment to provide better user experience.

\begin{figure}[t]
  \centering
\includegraphics[width=8cm,height=11cm]{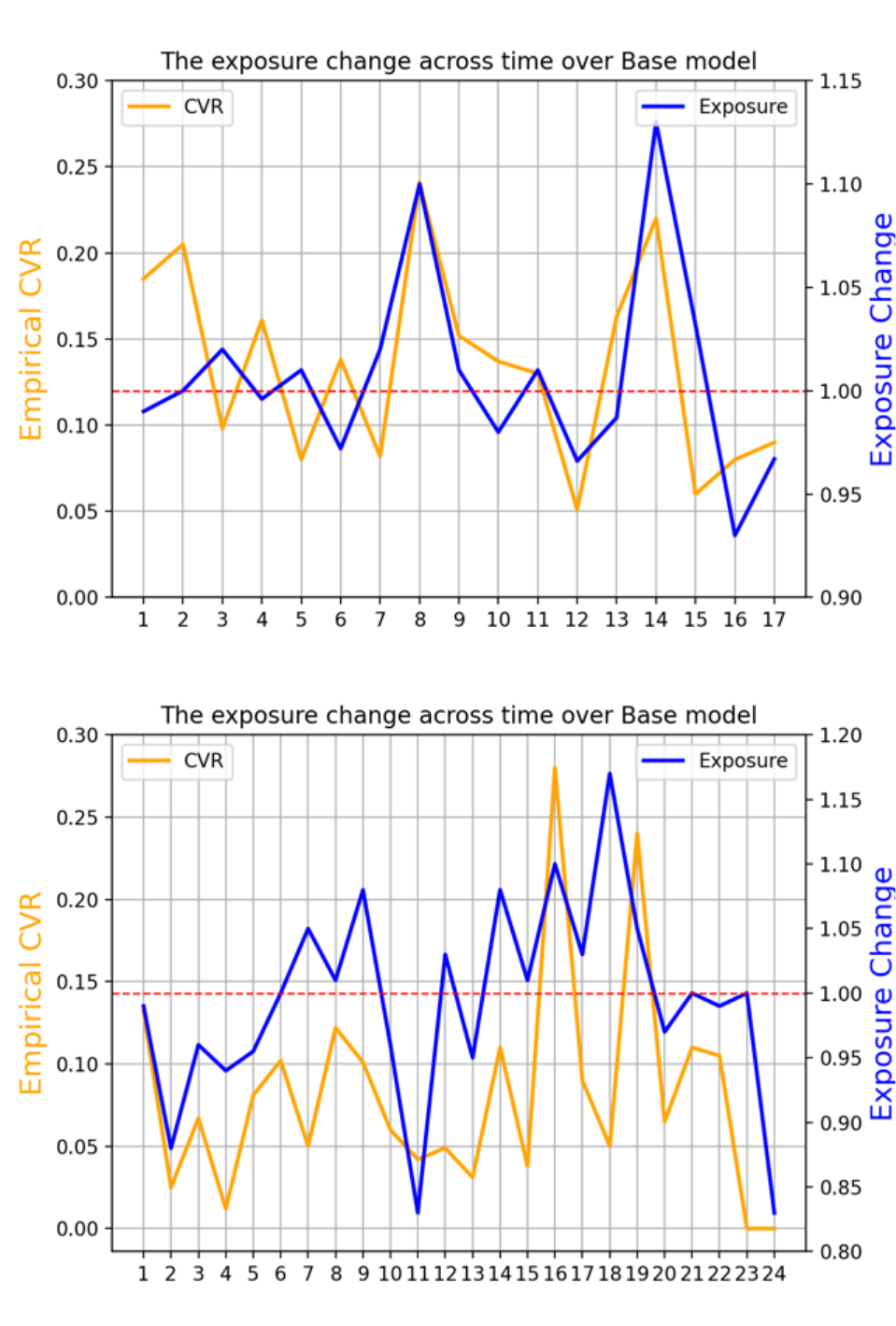}
  \caption{Live-streaming exposure trend across time.}
  \label{exposure}
\end{figure}

\subsection{Discussion of the Generative Accuracy (RQ3)}
As shown in Table~\ref{acc_abl1} and Table~\ref{acc_abl2}, we analyze the influence of the accuracy of statistic sequence and product sequence on the ranking model behavior. Specifically, we use different basic forecast method to forecast the future statistic sequence and product sequence, and use these features to replace the one forecast by Transformer. For statistic sequence, the basic forecast methods include using the average value or the latest value as the forecast result. For product sequence, the basis forecast methods include using the product with the most counts or using the latest product as the forecast result. We compare the forecast error and the model performance between the basic forecast methods and our method.
For both the statistic sequence and product sequence, our method has a lower forecast error. Meanwhile, the model performance increases as the forecast error decreases, indicating that the forecast performance is relevant to the ranking model performance.

\subsection{Discussion of the Statistic Sequences (RQ4)}
We also conduct experiments to analyze the influence of statistic Setting in Table~\ref{importance_abl}. The sequences include many types, each type may have a difference influence to the CTR or CVR prediction. We conduct experiments to analyze it, adding each type of features in CTR and CVR models. The experimental results show that CTR and CVR exhibit a different react to different kinds of features. For CTR prediction, the out-live features are more important. For CVR prediction, the interaction and in-live features are more important.
In addition, we conduct experiments on how forecasting steps influence the final result in Table~\ref{step_abl}. As the forecast steps increase, the forecast accuracy decreases, and the help of the ranking model is also decreasing. This suggests that forecast accuracy is important, and it is difficult to forecast the steps far ahead accurately.
In LiveForesighter, we only predict the next 3 steps for those statistic sequence in training and inference.

\subsection{Discussion of the Online Effect (RQ5)}
This section explores the LiveForesighter how to influence our system, as discussed in Introduction, our goal is to enhance the live-streaming distribution when at its `high-light' moment.
To valid our LiveForesighter effectiveness as expected, we show two online-shopping live-streaming cases in Figure~\ref{exposure} that describe live-streaming empirical CVR and the real exposure change across time (each time point represents 3 minutes).
Particularly, the exposure means a successful distribution to our users, and the empirical CVR rate denotes the proportion of actual purchases by users relative to exposure.
Regarding the Figure~\ref{exposure}, compared with base model exposure (in red dotted line), we could observe that when the CVR rate of a live-streaming at a high-level, our system can give more exposure to the live-streaming, and vice versa.
Such observation indicates our LiveForesighter is a trustworthy way to identify high-quality live-streaming in real-time setting.

\section{Related works}

\subsection{Live-Streaming Recommendation}
Live-streaming is a new media that has emerged in recent years, which allows users and authors to communicate and interact in real time, and has widely attracted the public and has accumulated a large user group. 
In recent years, live-streaming has gradually become professionalized, and a number of high-quality platforms have emerged, such as Kuaishou (Chatting, Shopping), Twitch (Gaming) and so on.
In industry, the Twitch proposed a representative work the LiveRec~\cite{liverec} to consider the users repeat watching pattern and utilize the attention mechanism to capture the watching frequency.
Later, considering users watch different live-streaming times of spans a large range, the Twitch proposed a re-weighting technique~\cite{chen2022weighing} to increase long-view samples weight adaptively.
To model the user-item relationship more fine-grained, the MMBee~\cite{deng2024mmbee} introduces the user-author graph to capture higher-order user-item connection information via different meta-paths.
With the wave of multi-modal learning, the MTA~\cite{xi2023multimodal} and ContentCTR~\cite{contentctr} are proposed to model live-streaming content as an additional feature.
Besides, to fulfill user interests, there are some works eLiveRec~\cite{eliverec} to consider the users other interaction behaviours in product/short-video domain.
In data-streaming designing for live-streaming, the Cross\&Moment~\cite{cao2024moment} describes the fast-slow window and 30s-window paradigm to train model in a real-time manner.
Compared with them, our LiveForesighter is from a new perspective to enhance live-streaming recommendation via generate future information, and further providing solutions to alleviate the following challenges:
(1) \textit{for a live-streaming, what moment is the best timing to distribute?}
(2) \textit{in addition to the current moment, does the future content also meet users interests?}
With the live-streaming side user-behaviour statistic and product sequences, our LiveForesighter achieves a disentanglement generative paradigm to achieve more smart live-streaming RecSys.

\subsection{Generative Models in Recommendation}
Generative models have been widely studied in recent years with the explosive development of natural language processing.
The elaborate works of generative models are GPTs~\cite{achiam2023gpt, bai2023qwen}, which stack large-scale Transformer architectures to learn the natural language corpus to generate next token.
With the success of GPTs, there are many fields have been revolutionized by generative models, such as the visual modeling KLing~\cite{tian2024videotetris}, time series forecasting Timer~\cite{liutimer}, speech modeling FunAudioLLM~\cite{an2024funaudiollm} and so on. 

For the recommendation area, the revolution of large generative models is gradually beginning.
At the retrieval stage, the KuaiFormer~\cite{kuaiformer} employs the Transformer as backbone to extract multiple user interests representation.
Meanwhile, the TIGER~\cite{tiger} and LIGER~\cite{yang2024unifying} utilize the LLM-generated Semantic ID to model user historical sequence with beam-search technique to generate the thousands of item candidate Semantic ID for more comprehensive retrieval.
The NoteLLM~\cite{zhang2024notellm} and QARM~\cite{luo2024qarm} utilize the item-item relationship to finetune the LLM to provide more friendly representation for recommendation model.
At the ranking stage, the HSTU~\cite{zhai2024actions} devise a generative multi-layer Transformer framework to replace wide-used ranking model to rank user and multiple items at same time.
Besides, MARM~\cite{lv2024marm} applies the KV-cache technique to achieve a low-computation paradigm, to scale the Transformers in ranking model efficiently.
The HLLM~\cite{chen2024hllm} utilizes the pre-trained language model to act as ranking model, which uses the multi-modal singals to represent items representation and user interests.
Compared with them, our LiveForesighter also follows the generative paradigm but with distinct motivations to predict future live-streaming information with Transformers.

\section{Conclusion}
To tackle the in-depth challenge of live-streaming services: \textit{How to discover the live-streamings that the contents user is interested in at the current moment, and further a period of time in the future?}
In this paper, we propose a novel generative model, LiveForesighter to inject the future information to enhance live-streaming recommendation.
To find the right live-streaming distribution timing, we utilize the live-streaming side user behaviours statistic sequences to monitor the real-time content quality to detect whether the live-streaming is in its high-light moment.
To detect the future live-streaming content is also meet user interests, we apply the product sequence to generate future next products category, to encourage user watching for a long-time.
Empirically experimental results on Kuaishou’s online-shopping and talent-show live-streaming scenarios demonstrate the effectiveness of our generative model LiveForesighter.
Further, detailed analyses describe the comprehensive effect of LiveForesighter to our online recommendation system.
In the future, we will investigate the potential of pre-trained multi-modal large language model to predict more fine-grained information to enhance live-streaming recommendation.

\balance
\bibliographystyle{ACM-Reference-Format}
\bibliography{sample-base.bib}
\end{document}